\documentclass[12pt]{article}
\usepackage{epsfig}
\def\be{\begin{equation}}
\def\ee{\end{equation}}
\def\bea{\begin{eqnarray}}
\def\eea{\end{eqnarray}}
\usepackage{graphicx}

\catcode`\@=11
\def\lsim{\mathrel{\mathpalette\@versim<}}
\def\gsim{\mathrel{\mathpalette\@versim>}}
\def\@versim#1#2{\vcenter{\offinterlineskip
\ialign{$\m@th#1\hfil##\hfil$\crcr#2\crcr\sim\crcr } }}
\catcode`\@=12
\usepackage{axodraw}

\catcode`@=12
\begin{document}
\thispagestyle{empty}
\begin{flushright}
UCRHEP-T470\\
July 2009\
\end{flushright}
\vspace{0.8in}
\begin{center}
{\LARGE \bf Seesaw Options for Three Neutrinos\\}
\vspace{1.2in}
{\bf Xiao-Gang He$^a$ and Ernest Ma$^b$\\}
\vspace{0.2in}
{\sl $^a$ Department of Physics and Center for Theoretical Sciences,\\
National Taiwan University, Taipei 10617, Taiwan\\}
\vspace{0.1in}
{\sl $^b$ Department of Physics and Astronomy, University of California,\\
Riverside, California 92521, USA\\}
\end{center}
\vspace{1.2in}
\begin{abstract}\
The seesaw mechanism for three neutrinos is discussed, clarifying the
situation where the seesaw texture results in three approximately zero
mass eigenvalues.  The true underlying mechanism is shown to be just
the $inverse$ (or $linear$) seesaw, which explains why there could be
large mixing. However, these zeroes cannot occur naturally, unless there
is a conserved symmetry, i.e. lepton number $L$, either global or gauged,
which is softly or spontaneously broken at the TeV scale.  We discuss in
particular the case where the three heavy singlet neutrinos have $L=3,-2,-1$.
\end{abstract}

\newpage
\baselineskip 24pt
In the famous canonical seesaw mechanism, the standard model
(SM) of particle interactions is implemented with a heavy singlet
``right-handed'' neutrino $N_R$ per family, so that the otherwise massless
left-handed neutrino $\nu_L$ gets a mass from diagonalizing the $2 \times 2$
mass matrix spanning
($\bar{\nu}_L, N_R$):
\begin{equation}
{\cal M}_{\nu,N} = \pmatrix{0 & m_D \cr m_D & m_N},
\end{equation}
resulting in
\begin{equation}
m_\nu \simeq {-m_D^2 \over m_N},
\end{equation}
with mixing between $\nu_L$ and $N_R$ given by
\begin{equation}
\tan \theta  \simeq {m_D \over m_N} \simeq \sqrt{|m_\nu/m_N|}.
\end{equation}
As a result, the $3 \times 3$ mixing matrix linking the 3 light neutrinos to
the 3 charged leptons cannot be exactly unitary.  However, for $m_\nu \sim 1$
eV and $m_N \sim 1$ TeV, this violation of unitarity is of order $10^{-6}$,
which is much too small to be observed.

Suppose the $6 \times 6$ mass matrix spanning $\nu_{1,2,3}$ and $N_{1,2,3}$
has three zero mass eigenvalues, without requiring $m_D = 0$ identically
\cite{bw90}, then it has been pointed out that the addition of small
perturbations to this texture will result in acceptably small neutrino masses
as well as possible large mixing \cite{Pilaftsis:1991ug,p05,ks07,h09} between $\nu_{1,2,3}$
and $N_{1,2,3}$, in contrast to the case of only one family.  It this paper,
we will discuss what this really means, and show that the underlying mechanism
for the origin of this large mixing is just the $inverse$ seesaw
\cite{ww83,mv86,m87} with a conserved symmetry, i.e. lepton number $L$,
which may be global (and softly or spontaneously broken) or gauged (and
spontaneously broken).  We will implement this idea with a specific model
with $L=3,-2,-1$ for $N_{1,2,3}$.

For simplicity, consider first two families.  It has been argued that large
mixing between $(\nu_1,\nu_2)$ and $(N_1,N_2)$ may occur if the Dirac mass
matrix linking them is of the form
\begin{equation}
{\cal M}_D = \pmatrix{a_1 b_1 & a_1 b_2 \cr a_2 b_1 & a_2 b_2},
\end{equation}
in the basis where
\begin{equation}
{\cal M}_N = \pmatrix{M'_1 & 0 \cr 0 & M'_2}.
\end{equation}
In that case, the arbitrary imposed condition
\begin{equation}
{b_1^2 \over M'_1} + {b_2^2 \over M'_2} = 0
\end{equation}
renders all two light neutrinos massless, without requiring ${\cal M}_D = 0$.
To understand what this really means, first note that the determinant of
${\cal M}_D$ is zero, hence there is only one nonzero eigenvalue.  Then
consider the most general $4 \times 4$ mass matrix spanning
$(\nu_1,\nu_2,N_1,N_2)$ in the basis where ${\cal M}_D$ is diagonal, i.e.
\begin{equation}
{\cal M}_{\nu,N} = \pmatrix{0 & 0 & m_1 & 0 \cr 0 & 0 & 0 & m_2 \cr m_1 & 0
& M_1 & M_3 \cr 0 & m_2 & M_3 & M_2}.
\end{equation}
Rotating the nondiagonal ${\cal M}_D$ of Eq.~(4) on the left with
$\tan \theta_L = a_1/a_2$ by the matrix
\begin{equation}
{\cal U}_L^\dagger = \pmatrix{\cos \theta_L & -\sin \theta_L \cr \sin \theta_L &
\cos \theta_L},
\end{equation}
and on the right with $\tan \theta_R = b_1/b_2$ by the matrix
\begin{equation}
{\cal U}_R = \pmatrix{\cos \theta_R & \sin \theta_R \cr -\sin \theta_R &
\cos \theta_R},
\end{equation}
the texture hypothesis is equivalent to setting
\begin{eqnarray}
&& m_1 = 0, ~~~ m_2 = \sqrt{a_1^2+a_2^2} \sqrt{b_1^2+b_2^2}, \\
&& M_1 = \cos^2 \theta_R M'_1 + \sin^2 \theta_R M'_2 = 0, \\
&& M_2 = \sin^2 \theta_R M'_1 + \cos^2 \theta_R M'_2 =
(1-\tan^2 \theta_R) M'_2, \\
&& M_3 = \sin \theta_R \cos \theta_R (M'_1-M'_2) = -\tan \theta_R M'_2.
\end{eqnarray}
It is then
clear that $\nu_1$ and the linear combination $\nu'_2 = (M_3 \nu_2 - m_2 N_1)
/\sqrt{M_3^2+m_2^2}$ are massless.  Once small perturbations are added, i.e.
$0 \neq m_1 << m_2$ and $0 \neq M_1 << M_{2,3}$, $\nu'_2$ gets a small mass
proportional to $M_1$ given by $(m_2^2/M_3^2)M_1$ through the inverse seesaw,
and the possibly large $\nu_2-N_1$ mixing remains.  The complete reduced
$2 \times 2$ mass matrix spanning $\nu_1$ and $\nu'_2$ is given by
\begin{equation}
{\cal M}_\nu \simeq \pmatrix{ m_1^2 M_2/M_3^2 & -m_1 m_2/M_3 \cr -m_1 m_2/M_3
& m_2^2 M_1/M_3^2}.
\end{equation}
Since $M_2 \sim M_3$ in this hypothesis, the (1,1) entry is a canonical seesaw,
whereas the (2,2) entry is an inverse seesaw.  The (1,2) or (2,1) entry is
known as the linear seesaw \cite{mrv05}, but it is equivalent to the inverse
seesaw, as explained in Ref.~\cite{m09-1}.  Note first that if $m_1=0$, then
only $\nu'_2$ gets a small mass (because $M_1$ is small) through the inverse
seesaw.  If $M_1=0$, then since $m_1 M_2/M_3 << m_2$ is assumed in such a
texture scenario, the two neutrinos are pseudo-Dirac partners and are
nearly degenerate in mass.  If $m_1 \neq 0$ and $M_1 \neq 0$, then it is
possible to have a solution where the (1,1) entry is negligible and the
other entries are comparable.

It has been argued that such a texture is protected by chiral symmetry.
Whereas this may be correct for $\nu_1$, it is obviously not true for $\nu'_2$
because $\nu_2$ couples to $N_2$, and $N_2$ has a nonzero Majorana mass, i.e.
$M_2$.  The one-loop diagram connecting $\nu_2$ to itself through $N_2$ and
the SM Higgs boson is infinite and there is no corresponding diagram from
$N_1$ to cancel it.  Thus the Majorana mass of $\nu'_2$ has an infinite
correction and cannot be zero naturally. The texture idea alone has no
support in terms of a symmetry.

On the other hand, if $M_2 = 0$, then a conserved lepton number $L$ can be
defined, with $L=1$ for $N_1$ and $L=-1$ for $N_2$.  If small $M_{1,2}$ and
$m_1$ are now added, thus breaking $L$ to $(-1)^L$, Eq.~(14) will be obtained
with a very small (1,1) entry.

To maintain Eq.~(7) with $m_1=M_1=0$ and $M_2 \sim M_3$, the lepton-number
global symmetry has to be redefined, with for example $L=3,-1$ for
$N_{1,2}$.  In that case, the addition of the standard Higgs doublet $\Phi_1
= (\phi_1^+,\phi_1^0)$ with $L=0$ will link $\nu_2$ with $N_2$ to obtain $m_2$,
whereas a Higgs singlet $\chi_2$ with $L=2$ will supply $N_2$ with the Majorana
mass $M_2$, and its complex conjugate $\chi_2^\dagger$ will link $N_1$ with $N_2$
to obtain $M_3$.  The absence of a Higgs singlet with $L=6$ will forbid a
Majorana mass $M_1$ for $N_1$ at tree level, but it will be induced by
the mass splitting of Re($\chi_2$) and Im($\chi_2$) in one loop after the
breaking of $U(1)_L$, as shown in Fig.~1.
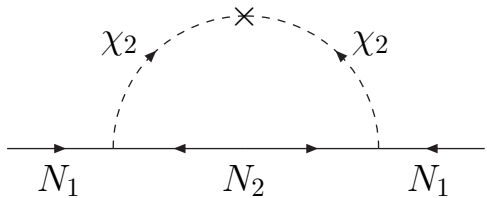
\begin{figure}[htb]
\begin{center}
\begin{picture}(360,90)(0,0)
\ArrowLine(90,10)(130,10)
\ArrowLine(180,10)(130,10)
\ArrowLine(180,10)(230,10)
\ArrowLine(270,10)(230,10)
\DashArrowArcn(180,10)(50,180,90)3
\DashArrowArc(180,10)(50,0,90)3

\Text(110,-2)[]{\large$N_1$}
\Text(250,-2)[]{\large$N_1$}
\Text(180,-2)[]{\large$N_2$}
\Text(133,50)[]{\large$\chi_2$}
\Text(228,50)[]{\large$\chi_2$}
\Text(180,60)[]{\large$\times$}
\end{picture}
\end{center}
\caption{One-loop generation of $M_1$.}
\end{figure}
This diagram is finite because of the cancellation between Re($\chi_2$) and
Im($\chi_2$).  If $U(1)_L$ is spontaneously broken, then Im($\chi_2$) is a
massless Goldstone boson, i.e. the singlet majoron.  If $U(1)_L$ is
explicitly broken but only softly, with the addition of the term
$\mu^2 \chi_2^2 + H.c.$ for example, then Im($\chi_2$) is massive.

Consider now the most general $6 \times 6$ mass matrix spanning
$(\nu_{1,2,3},N_{1,2,3})$:
\begin{equation}
{\cal M}_{\nu,N} = \pmatrix{0 & 0 & 0 & m_1 & 0 & 0 \cr 0 & 0 & 0 & 0 & m_2
& 0 \cr 0 & 0 & 0 & 0 & 0 & m_3 \cr m_1 & 0 & 0 & M_1 & M_4 & M_5 \cr 0 & m_2
& 0 & M_4 & M_2 & M_6 \cr 0 & 0 & m_3 & M_5 & M_6 & M_3}.
\end{equation}
The texture hypothesis is equivalent to $m_1 = m_2 = 0$ and $M_1
= M_4 = 0$.  Again it is clear that there could be large mixing between
$\nu_3$ and $N_1$, but there is no symmetry which enforces it.
Consider now lepton number with $L=1$ for
$N_{1,2}$ and $L=-1$ for $N_3$, then $M_1=M_2=M_3=M_4=0$ and the linear
combination $(M_6 N_1 - M_5 N_2)/\sqrt{M_5^2+M_6^2}$ is massless.
Once small perturbations are added, this becomes a scenario for four light
neutrinos (three active and one sterile).  Suppose $N_1$ has $L=1$, $N_2$
has $L=0$, $N_3$ has $L=-1$, then $M_1=M_3=M_4=M_6=0$.  In this case, $N_2$
has mass $M_2$, and there are exactly three massless neutrinos.

To maintain the seesaw texture $m_{1,2}=0$, $m_3 \neq 0$, $M_{1,4}=0$, and
$M_{2,3,5,6} \neq 0$, the lepton-number global symmetry must again be
redefined.  Let $\nu_{1,2,3}$ have $L=1$ as usual, and $N_{1,2,3}$ have
$L=3,-2,-1$ respectively.  Let there again be a Higgs doublet $\Phi_1$
with $L=0$ and now three Higgs singlets $\chi_{2,3,4}$ with $L=2,3,4$.
Then $M_3$ comes from $\langle \chi_2 \rangle$, $M_5$ from $\langle
\chi_2^\dagger \rangle$, $M_6$ from $\langle \chi_3 \rangle$, and $M_2$
from $\langle \chi_4 \rangle$.  The three massless eigenstates are
\begin{equation}
\nu_1, ~~~ \nu_2, ~~~ \nu'_3 = {M_5 \nu_3 - m_3 N_1 \over \sqrt{M_5^2 + m_3^2}},
\end{equation}
showing explicitly how $\nu_3-N_1$ mixing can be large even if all neutrinos
are massless.  The analog of Fig.~1 now applies to $M_1$ and $M_4$, both of
which obtain one-loop finite masses, resulting in an inverse seesaw mass for
$\nu'_3$, i.e. $M_1 m_3^2/M_5^2$.  As for $\nu_{1,2}$ masses, we need extra
Higgs doublets.  Consider the minimal case of a second Higgs doublet
$\Phi_2 = (\phi_2^+,\phi_2^0)$ with $L=1$.  It couples $\nu_{1,2,3}$ to $N_2$.
By redefining $\nu_{1,2}$, we consider only the couplings to $\nu_{2,3}$,
resulting in the masses $m_{22}$ and $m_{32}$.  Thus $\nu_1$ remains
massless and the reduced $2 \times 2$ mass matrix spanning $\nu_2$ and
$\nu'_3$ is given by
\begin{equation}
\pmatrix{-m_{22}^2/M_2 & -(m_{22}/M_2)[m_{32} + m_3 (M_1 M_6 -
M_4 M_5)/M_5^2] \cr -(m_{22}/M_2)[m_{32} + m_3 (M_1 M_6 - M_4 M_5)/M_5^2]
& M_1 m_3^2/M_5^2}.
\end{equation}
In the above, let $m_{22} \sim m_{32} \sim M_1 \sim M_4 \sim 1$ MeV,
$m_3 \sim 1$ GeV, and $M_2 \sim M_3 \sim M_5 \sim M_6 \sim 1$ TeV, then
all entries are of order 1 eV, and suitable for a realistic neutrino mass
matrix, allowing for both normal and inverse hierarchies.

Another minimal case is to add a second Higgs doublet $\Phi_2 =
(\phi_2^+,\phi_2^0)$ with $L=-4$ instead.  Now we have $m_{21}$ and $m_{31}$
instead, and the reduced $2 \times 2$ mass matrix ${\cal M}_\nu$ spanning \
$\nu_{2}$ and $\nu'_3$ is given by
\begin{eqnarray}
\pmatrix{ -m_{21}^2 (M_6^2-M_2M_3)/M_2 M_5^2 & -m_{21}m_{3}/M_5 \cr -m_{21} m_3
/M_5 & M_1 m_3^2/M_5^2 -2m_{31} m_{3}/M_5}.
\end{eqnarray}
This structure is different from Eq.~(17) but similar to Eq.~(14).  The
off-diagonal entries could be much bigger than the diagonal ones (if $m_{31}
<< m_{21}$), thereby allowing for two nearly degenerate neutrino masses,
which is perfect for understanding an inverse hierarchy, where the mass
splitting responsible for solar neutrino oscillations is small compared
to the neutrino masses themselves.  On the other hand, if $m_{21} << m_{31}$,
normal hierachy is also possible.  Once $\nu_{2,3}$ are massive, $\nu_1$ will
acquire a nonzero mass through the exchange of two $W$ bosons \cite{bm88},
but this contribution is negligible.

Consider now the Higgs potential of $\Phi_{1,2}$ (with $L=-4$ for $\Phi_2$)
and $\chi_{2,3,4}$, invariant under $U(1)_L$:
\begin{eqnarray}
V &=&  \sum_{i=1,2} \mu_i^2 \Phi_i^\dagger \Phi_i + \sum_{i=2,3,4} m_i^2
\chi_i^\dagger \chi_i + {1 \over 2} \sum_{i,j=1,2} \lambda_{ij}
(\Phi_i^\dagger \Phi_i)(\Phi_j^\dagger \Phi_j) + \lambda'_{12}
(\Phi_1^\dagger \Phi_2)(\Phi_2^\dagger \Phi_1) \nonumber \\
&+& {1 \over 2} \sum_{i,j=2,3,4} f_{ij} (\chi_i^\dagger \chi_i)
(\chi_j^\dagger \chi_j) +  \sum_{i=1,2,j=2,3,4} h_{ij} (\Phi_i^\dagger \Phi_i)
(\chi_j^\dagger \chi_j) \nonumber \\ &+& [\mu_{124} \Phi_1^\dagger \Phi_2
\chi_4 + m_{224} \chi_2^2 \chi_4^\dagger + h_{122} \Phi_1^\dagger \Phi_2
\chi_2^2 + f_{234} \chi_2^\dagger \chi_3^2 \chi_4^\dagger + H.c.]
\end{eqnarray}
To have $\langle \phi_2^0 \rangle << \langle \phi_1^0 \rangle$, the
couplings $\mu_{124}$ and $h_{122}$ must be chosen to be very small, and
$\mu_2^2$ positive and large \cite{m01}.  It may be argued that $\mu_{124}$
and $h_{122}$ are naturally small because if they were zero, then $V$ would
have an extra global U(1) symmetry, in addition to $U(1)_L$.  As it is,  
there is no extra global U(1), but the spontaneous breaking of $U(1)_L$ 
does result in a massless Goldstone boson, the singlet majoron.  To
avoid this complication, soft explicit $U(1)_L$ breaking terms, such as
$\mu^2_{12} \Phi_1^\dagger \Phi_2 + H.c.$, could be added.  If $L=1$ is chosen
for $\Phi_2$, then the $\mu_{124}$ and $h_{122}$ terms of Eq.~(19) are replaced
by $h_{1223} \Phi_1^\dagger \Phi_2 \chi_2 \chi_3^\dagger + h_{1234} \Phi_1^\dagger
\Phi_2 \chi_3 \chi_4^\dagger + H.c.$ and everything works just as well.

Lepton number may also be considered as a discrete symmetry, in which case
$Z_7$ works for the case $L=-4$ (which is equivalent to $L=3$) for $\Phi_2$.
Now $\chi_3$ is equivalent to $\chi_4^\dagger$ and should be eliminated.  The
new term $\chi_2 \chi_4^3$ would now appear, by which the massless majoron is
eliminated.

An alternative is to gauge the global $U(1)_L$ symmetry, using either
$U(1)_{B-L}$ or $U(1)_\chi$ from the decomposition of $SO(10) \to SU(5)
\times U(1)_\chi$, where $Q_\chi = 5(B-L)-4Y$.  The same seesaw texture
may be maintained using exactly the same lepton number assignments.  The
difference is that there can be no soft symmetry breaking terms and the
extra anomalies generated by $N_{1,2}$ should be offset, for example, by
three pairs of singlets with $L=1,-2$, belonging to a separate (odd $Z_2$)
sector.

Consider now specifically $U(1)_\chi$ \cite{m09-2}. Since $\Phi_2$ has
nonzero $Q_\chi$, its vacuum expectation value $\langle \phi_2^0 \rangle$
contributes to $Z-Z'_\chi$ mixing which is known to be very small
\cite{elmr09}.  This fits perfectly into our scenario because $m_{22},m_{32}$
and $m_{21},m_{31}$ are also proportional to $\langle \phi_2^0 \rangle$, and
have been chosen to be small for neutrino masses.  Constraints on $Z'_\chi$
then come mainly from its direct search at the Tevatron and the anomalous
$g-2$ value of the muon.  The present best direct lower limit for the mass
$M_{Z'_\chi}$ is 822 GeV \cite{cdf07}.  Using this bound, the muon $g-2$
constraint is easily satisfied as well.

If $M_{Z'_\chi}$ is not too much larger than the present lower limit, it can
be produced at the Large Hadron Collider (LHC), due to start taking data
soon this year.  Since $Z'_\chi$ couples to SM particles with different
$U(1)_\chi$ charges: 1, $-1$ and 3 for left-handed quark doublets, right-handed
$up$ and $down$ quark singlets; $-3$ and $-1$ for left-handed and right-handed
charged leptons, the forward-backward asymmetries in $b \bar b$ and charged
lepton-pair production will deviate from pure $Z$ exchange.  This may provide
a signal of new physics beyond the SM.

The $Z'_\chi$ boson can also decay into final states containing the heavy
singlet neutrinos.  If $M_{Z'_\chi} > 2m_N$, then $Z'_\chi$ will decay into
$N \bar N$ with subsequent decays $N \to l^- W^+,\;\nuƒËZ$ and $\bar N \to¨
l^+ W^-, \;\bar \nu Z$, etc.  Depending on which $N$ is the lightest and
which ones are produced, the signature may be different. If $N$ is Majorana,
which is possible for $N_2$,  then the final decay products of $Z'_\chi$
can have both $e^\pm e^\mp W^\mp W^\pm$ and $e^\pm e^\pm W^\mp W^\mp$.  If
$N$ is from one of the linear combinations of $N_{1,3}$, and $M_3$ is much
smaller than $M_5$, the mass eigenstate can be a Dirac
particle paired from $N_1$ and $N^c_3$.  If so, then the final product will
have just $l^\pm l^\mp W^\mp W^\pm$.  If the mass eigenstates have large
Majorana components, i.e. $M_3 \sim M_5$, the final products also have
significant $l^\pm l^\pm W^\mp W^\mp$ event rates.

There is another potentially large decay channel involving a single heavy
neutrino, i.e. $Z'_\chi \to \nu N$, because large mixing between light
and heavy neutrinos is possible.  This will be the dominant
channel producing heavy neutrinos from $Z'_\chi$ decay for $M_{Z'_\chi}$
in the range $m_N < M_{Z'_\chi} < 2 m_N$.

It is obvious that the best way to verify the seesaw mechanism is to produce
the heavy singlet neutrinos.  The presence of $Z'_\chi$ allows for this to
happen much more easily than in models without it.  In the latter type of
models, the production of $N$ is through the single production channel,
$q \bar q \to Z \to \nu N$ and $q\bar q' \to W \to l N$ with the subsequent
decay of $N$ into $l W$.  This mechanism is not completely negligible because
the texture hypothesis allows for large mixing between light and heavy
neutrinos.  It has been shown \cite{seesaw1-collider} that $m_N$ up to a
hundred GeV may be probed at the LHC.  The detection of such a single $N$ can
provide useful information on the texture hypothesis discussed in this
work. By looking at the decaying vertex of $N$, one can also estimate the
size of mixing between light and heavy neutrinos. In the canonical seesaw
case, mixing of order $(m_\nu/m_N)^{1/2}$ leads to a very small decay width
for $N$.  Although it is not stable enough to escape the detector, it will
produce a displaced vertex.  This will not be the case for the large mixing
being considered here.

With $Z'_\chi$, it is possible to produce $N$ in pairs through, $q \bar q
\to Z' \to N \bar N$, if kinematically allowed.  The final states to be
analyzed are $l^\pm l^\mp W^\pm W^\mp$ and $l^\pm l^\pm W^\mp W^\mp$.  The
situation is similar to that of the Type III seesaw model \cite{seesaw3}
where the charged partner $E^\pm$ of the neutral heavy neutrino in the
$SU(2)_L$ triplet is analyzed using $q\bar q \to Z \to E^+ E^-$ with $E$
subsequently decaying into $l Z$ \cite{seesaw3-collider}. There $m_E$ up
to a TeV can be probed.  In this model, however, the cross section will be
smaller because the heavier $Z'_\chi$ is mediating the interaction, except
of course if the production is at the $Z'_\chi$ resonance, which is the main
advantage of having $U(1)_\chi$.  A possible scenario is thus the discovery
of $Z'_\chi$ at the LHC and from a detailed study of its decay products,
the heavy neutrino states are also discovered with the information
necessary to reconstruct the appropriate seesaw texture.

To conclude, we have studied the seesaw mechanism for three neutrinos,
clarifying the situation where the texture of the $6 \times 6$ mass matrix
results in three approximately zero mass eigenvalues.  The true underlying
mechanism is shown to be just the $inverse$ (or $linear$) seesaw, which
explains why there could be large mixing.  However, these zeroes cannot
occur naturally, unless there is a conserved symmetry, i.e. lepton number
$L$, either global, discrete or gauged, which is softly or spontaneously
broken at the TeV scale.  We discuss in particular a case where the heavy
singlet neutrinos have $L=3,-2,-1$.  To support the texture hypothesis,
Higgs singlets must be added, and the zeros of the $3 \times 3$ mass matrix
of the heavy singlet neutrinos at tree level are shown to be nonzero in
one loop. The lepton symmetry may also be gauged, thereby predicting a
$Z'$ boson which would facilitate the discovery of the heavy singlet
neutrinos at the LHC.

This work was supported in part by the Taiwan NSC and NCTS, and the
U.~S.~Department of Energy under Grant No.~DE-FG03-94ER40837.

\bibliographystyle{unsrt}

\end{document}